\documentclass[a4paper,aps,preprint]{revtex4-1}
\usepackage{graphicx}

\begin{document}


\title{Quantum Dynamics in Classical Spin Baths}

\author{Alessandro Sergi}
\email{sergi@ukzn.ac.za}

\affiliation{
School of Chemistry and Physics, University of KwaZulu-Natal, Pietermaritzburg,
Private Bag X01 Scottsville, 3209 Pietermaritzburg\\
and \\
National Institute for Theoretical Physics (NITheP), KwaZulu-Natal, South Africa}

\begin{abstract}
A formalism for studying the dynamics of quantum systems embedded
in classical spin baths is introduced.
The theory is based on generalized antisymmetric brackets
and predicts the presence of open-path off-diagonal
geometric phases in the evolution of the density matrix.
The weak coupling limit of the equation can be integrated
by standard algorithms and provides
a non-Markovian approach to the computer simulation
of quantum systems in classical spin environments.
It is expected that the theory and numerical schemes presented here
have a wide applicability.
\end{abstract}

\maketitle

In recent years, dissipation and decoherence in quantum systems have come
to be considered more as resources to be exploited~\cite{plenio,plenio2,petr1}
than as simple nuisances to be avoided at all cost.
The theoretical study of both effects requires to couple the system of interest
to an environment~\cite{petruccione}.
This latter can be represented either by means of bosonic
degrees of freedom~\cite{leggett} or by spinors~\cite{prokofev,wang-shao,schnack}.
Full quantum theories of both system and environment are difficult to simulate
and usually one resorts to approached based on master equations~\cite{petruccione,ottinger-qm,ottinger-qm2,ottinger-qm3,ottinger-qm4,ottinger-qm5}.
However, there are many situations where the environment's coordinates
can be considered to follow the laws of classical mechanics.
When the such coordinates are canonically conjugate
momenta and positions, one approach to treat these situations
is provided by the quantum-classical Liouville equation~\cite{anderson,aleksandrov,prezhdo,qcl2,qcl3,qcl4,balescu2,mcquarrie,qcl5,qcl6,qcl8,qcl9,ilya,hanna,geva}.
Nevertheless, there are also many interesting cases when the bath 
is described in terms of classical spins, e.g., in order to model
complex molecules where magnetic effects
are important~\cite{schroeder,schroeder2}.
Such models can be studied by means of Monte Carlo methods
or by Molecular Dynamics simulations~\cite{allentildesley,frenkelsmit}.

The scope of this communication is to generalize
the quantum-classical Liouville equation~\cite{anderson,aleksandrov,prezhdo,qcl2,qcl3,qcl4,balescu2,mcquarrie,qcl5,qcl6,qcl8,qcl9,ilya,hanna,geva}
in order to study the
dynamics of quantum systems embedded in baths of classical spins.
This is naturally achieved within a quantum-classical theory
formulated by means of generalized 
antisymmetric brackets~\cite{b3,b4,b-silurante}.
It is worth noting that antisymmetric brackets have also been used
to formulate the statistical mechanics of classical systems with 
thermodynamic~\cite{b1,b2,sergi-pvg}
and holonomic constraints~\cite{holo1,holo2}.

The quantum-classical Liouville equation for
spin baths must be represented in some basis
to be of practical use and, in the regime of weak coupling,
the adiabatic basis is a good choice.
However, reasonable forms of the coupling between the quantum
subsystem and the classical spin bath
cause such a basis to be complex (whereas,
when the bath is described by canonical variables,
the adiabatic basis is real in the absence of magnetic fields).
Hence, an interesting prediction of the
quantum-classical Liouville equation for spin baths
is the presence of open-path~\cite{pati} 
geometric phases~\cite{berry,qphases,mead}
in the evolution of the off-diagonal elements~\cite{filipp,englman,manini}
of the density matrix.

In order to introduce the formalism,
one can consider a classical spin vector $\bf S$ 
whose energy is described by the Hamiltonian $H_{\rm SB}({\bf S})$,
and denote its components with $S_I$, $I=x,y,z$.
The generalization to systems with many spins is straightforward.
The equations of motion can be written in matrix form as
\begin{equation}
\dot{S}_I=\sum_J{\cal B}_{IJ}^{\rm S}\frac{\partial H_{\rm SB}}{\partial S_J}\;,
\label{eq:eqofm}
\end{equation}
where
\begin{equation}
\mbox{\boldmath $\cal B$}^{\rm S}
=\left[\begin{array}{ccc} 0   &  S_z & -S_y \\ 
                         -S_z &  0   &  S_x \\
                          S_y & -S_x &  0 \end{array}\right] 
\;.
\label{eq:bmat}
\end{equation}
The antisymmetric matrix $\mbox{\boldmath $\cal B$}^{\rm S}$ can also be written
in a compact way as
${\cal B}_{IJ}^{\rm S}=\epsilon_{IJK}S_K$,
where $\epsilon_{IJK}$ is the completely antisymmetric tensor
(or Levi-Civita symbol).
The equations of motion~(\ref{eq:eqofm})
preserve the Casimir $C_2={\bf S}\cdot {\bf S}$
for any arbitrary Hamiltonian $H_{\rm SB}({\bf S})$.
They also have have zero phase space compressibility
$\kappa=\sum_I\partial \dot{S}_I/\partial S_I =0$.
The classical equations of motion of the spin
can be written upon introducing the non-canonical bracket
\begin{equation}
\{A,B\}=\sum_{I,J}\frac{\partial A}{\partial S_I}{\cal B}_{IJ}^{\rm S}
\frac{\partial B}{\partial S_J}\;,
\label{eq:ncan-brack}
\end{equation}
where $A=A({\bf S})$ and $B=B({\bf S})$ are arbitrary functions
of the spin degrees of freedom,
in the form
\begin{equation}
\dot{S}_I=\{S_I,H_{\rm SB}\}\;.
\end{equation} 
At this stage, the quantum-classical theory can be introduced assuming that
the classical spin system interacts with 
a quantum system with a Hamiltonian $\hat{H}(\{\hat{\chi}\})$
through an interaction of the form $\hat{H}_{\rm c}(\{\hat{\chi}\},{\bf S})$
Hence, the total Hamiltonian operator of the quantum subsystem
in the classical spin bath can be written as
\begin{equation}
\hat{\cal H}({\bf S})=\hat{H}(\{\hat{\chi}\})
+\hat{H}_{\rm C}(\{\hat{\chi}\},{\bf S}) +H_{\rm SB}({\bf S})
\label{eq:tot-hamS}
\;.
\end{equation}
Given the quantum-classical density matrix $\hat{\rho}({\bf S},t)$,
the key step is postulating, in analogy with the case of
the quantum-classical Liouville
equation for canonical coordinates, the evolution of 
of the quantum  system in the classical spin bath 
by means of the equation
\begin{eqnarray}
\frac{\partial}{\partial t}\hat{\rho}({\bf S},t)
&=&-\frac{i}{\hbar}\left[\begin{array}{cc} \hat{\cal H}({\bf S}) & 
\hat{\rho}({\bf S},t)\end{array}\right]
\cdot\mbox{\boldmath$\cal D$}^{\rm S}\cdot
\left[\begin{array}{c}\hat{\cal H}({\bf S})\\ 
\hat{\rho}({\bf S},t)\end{array}\right]
\;,\nonumber\\
&=&-\frac{i}{\hbar}\left[\hat{\cal H}({\bf S}),\hat{\rho}({\bf S},t)
\right]_{\mbox{\tiny \boldmath$\cal D$}^{\rm S}}
\label{eq:pWS}
\end{eqnarray}
where
\begin{eqnarray}
&&\mbox{\boldmath$\cal D$}^{\rm S}
=\nonumber\\
&&\left[\begin{array}{cc} 0 & 1+\frac{i\hbar}{2}\sum_{I,J}
\frac{\overleftarrow{\partial}}{\partial S_I}{\cal B}_{IJ}^{\rm S}
\frac{\overrightarrow{\partial}}{\partial S_J}
\\
-1-\frac{i\hbar}{2}\sum_{I,J}
\frac{\overleftarrow{\partial}}{\partial S_I}{\cal B}_{IJ}^{\rm S}
\frac{\overrightarrow{\partial}}{\partial S_J}
& 0\end{array}\right]\;.
\nonumber\\
\end{eqnarray}
Equation~(\ref{eq:pWS}) is the quantum-classical Liouville equation
for quantum systems in classical spin baths.
Its proposal is the main result of this work.

The right hand side of Eq.~(\ref{eq:pWS}) defines what, in the case
of canonically conjugate phase space coordinates, is known as 
quantum-classical bracket~\cite{anderson,aleksandrov,prezhdo}
or non-Hamiltonian commutator~\cite{b3}.
The bracket couples the dynamics of the classical degrees of freedom
with that of the quantum operators taking into account
both the conservation of the energy and the quantum back-reaction.
Moreover, when there is no coupling, i.e., $\hat{H}_{\rm C}=0$,
the bracket makes the quantum system evolve in terms
of the standard commutator and the classical bath through
the Poisson bracket, which in the present case has the non-canonical form
given in Eq.~(\ref{eq:ncan-brack}).
The quantum-classical spin bracket in Eq.~(\ref{eq:pWS})
does not satisfy the Jacobi relation as its canonical
coordinate counterpart~\cite{anderson,aleksandrov,prezhdo}
and this, in turn,
leads to the lack of time-translation invariance
of the algebra defined in terms of the bracket itself~\cite{b3,nielsen}.
Less abstract consequence of such mathematical features are
that the coordinates of the spin bath, which are classical at time zero,
acquire quantum phases during the dynamics.
In practice, fast bath decoherence alleviates this problem
and, indeed, the canonical version of the
quantum-classical bracket, or of the quantum-classical Liouville equation,
is used for many applications in chemistry and 
physics~\cite{anderson,aleksandrov,prezhdo,qcl2,qcl3,qcl4,balescu2,mcquarrie,qcl5,qcl6,qcl8,qcl9,ilya,hanna,geva}.
Moreover, it is worth reminding that the non-Lie
(or, as they are also called, non-Hamiltonian) brackets,
with their lack of time translation invariance,
are also used to impose thermodynamical
(such as constant temperature and/or pressure)~\cite{b1,b2,sergi-pvg}
and holonomic constraints~\cite{holo1,holo2}
in classical molecular dynamics simulations~\cite{allentildesley,frenkelsmit}.


In order to represent the abstract Eq.~(\ref{eq:pWS}) the
quantum-classical Hamiltonian of Eq.~(\ref{eq:tot-hamS})
can be split as
\begin{equation}
\hat{\cal H}({\bf S})
=H_{\rm SB}({\bf S})+\hat{h}({\bf S})
\;.
\end{equation}
Accordingly, the adiabatic basis is defined by the eigenvalue equation
\begin{equation}
\hat{h}({\bf S})|\alpha;{\bf S}\rangle =E_{\alpha}({\bf S})|\alpha;{\bf S}\rangle\;.
\label{eq:adbasS}
\end{equation}
As one can see, the adiabatic basis in spin baths depends on all the non-canonical spin
coordinates $\bf S$, while in the canonical case it depends
only on the positions $R$ and not on the conjugate momenta $P$.
In such a basis, Eq.~(\ref{eq:pWS}) becomes
\begin{eqnarray}
\partial_t\rho_{\alpha\alpha'}
&=&
-i\omega_{\alpha\alpha'}\rho_{\alpha\alpha'}
- \sum_{I,J}{\cal B}_{IJ}^{\rm S}\left[ \frac{\partial H_{\rm SB}}{\partial S_J}
\langle\alpha|\frac{\partial\hat{\rho}}{\partial S_I}|\alpha'\rangle
\right.
\nonumber\\
&+&\left.\frac{1}{2}
\langle\alpha|\frac{\partial\hat{h}}{\partial S_I}
\frac{\partial\hat{\rho}}{\partial S_J}|\alpha'\rangle
-\frac{1}{2}
\langle\alpha|\frac{\partial\hat{\rho}}{\partial S_I}
\frac{\partial\hat{h}}{\partial S_J}|\alpha'\rangle
\right]
\nonumber\\
\end{eqnarray}
where we have used the antisymmetry of $\mbox{\boldmath$\cal B$}^{\rm S}$.
Defining the coupling vector
\begin{eqnarray}
d^I_{\sigma\alpha}&=&\langle\sigma|\frac{\overrightarrow{\partial}}{\partial S_I}|\alpha\rangle
\;
\end{eqnarray}
one finds the two identities
\begin{eqnarray}
\langle\alpha|\partial_I\hat{\rho}|\alpha'\rangle
&=&
\partial_I\rho_{\alpha\alpha'}
+ d^{I}_{\alpha\sigma}\rho_{\sigma\alpha'}
-\rho_{\alpha\sigma'} d^I_{\sigma'\alpha'}
\label{eq:drho}
\\
\langle\alpha|\frac{\partial\hat{h}}{\partial S_I}|\sigma\rangle
&=&
\frac{\partial h_{\alpha\sigma}}{\partial S_I}
-\Delta E_{\alpha\sigma} d^I_{\alpha\sigma}
\label{eq:dh}
\end{eqnarray}
where $\Delta E_{\alpha\sigma}=E_{\alpha}-E_{\sigma}$.
With the help of Eqs.~(\ref{eq:drho}) and~(\ref{eq:dh}),
the quantum-classical
Liouville equation for spin baths is written as
\begin{eqnarray}
\partial_t\rho_{\alpha\alpha'}
&=&\sum_{\beta\beta'}\left(-i\omega_{\alpha\alpha'}\delta_{\alpha\beta}\delta_{\alpha\alpha'}
-L_{\alpha\alpha'}\delta_{\alpha\beta}\delta_{\alpha\alpha'}
-J_{\alpha\alpha',\beta\beta'}
\right.\nonumber\\
&+&\left.{\cal S}_{\alpha\alpha',\beta\beta'}\right)
\rho_{\beta\beta'}\;.
\label{eq:qcsd-dyna}
\end{eqnarray}
In Eq.~(\ref{eq:qcsd-dyna})
the classical-like spin-Liouville operator
\begin{eqnarray}
L_{\alpha\alpha'}&=&
\sum_{I,J} {\cal B}_{IJ}^{\rm S}
\left( \frac{\partial H_{\rm SB}}{\partial S_J} \frac{\partial}{\partial S_I}
+\frac{1}{2} \frac{\partial E_{\alpha'}}{\partial S_J} \frac{\partial}{\partial S_I}
\right.\nonumber\\
&+&\left.
\frac{1}{2} \frac{\partial E_{\alpha}}{\partial S_J} \frac{\partial}{\partial S_I}
\right)
= \sum_{I,J}
{\cal B}_{IJ}^{\rm S}\frac{\partial H_{\alpha\alpha'}^{\rm S}}{\partial S_J} \frac{\partial}{\partial S_I}
\label{eq:Lspin}
\end{eqnarray}
has been defined.
The symbol $H_{\alpha\alpha'}^{\rm S}$ denotes
the average adiabatic surface Hamiltonian
\begin{equation}
H_{\alpha\alpha'}^{\rm S}=H_{\rm SB}+\frac{1}{2}\left(E_{\alpha}+E_{\alpha'}\right)
\;.
\end{equation}
The transition operator for the spin bath is given by
\begin{eqnarray}
J_{\alpha\alpha',\beta\beta'}
&=&
\sum_{I,J}{\cal B}_{IJ}^{\rm S}
\left[\frac{\partial H_{\rm SB}}{\partial S_J} d^{I}_{\alpha\beta}\delta_{\beta'\alpha'}
+\frac{1}{2}\Delta E_{\alpha\beta}d^{I}_{\alpha\beta}\frac{\partial}{\partial S_J}
\delta_{\alpha'\beta'}
\right.
\nonumber\\
&+& \left.\frac{\partial H_{\rm SB}}{\partial S_J} d^{I*}_{\alpha'\beta'} \delta_{\alpha\beta}
+\frac{1}{2}\Delta E_{\alpha'\beta'}d^{I*}_{\alpha'\beta'}\frac{\partial }{\partial S_J}
\delta_{\alpha\beta}
\right]
\;.
\nonumber\\
\label{eq:Jspin}
\end{eqnarray}
The operator in Eq.~(\ref{eq:Jspin}) goes to the usual jump operator~\cite{qcl6}
when canonical variables are considered. 
For the spin bath a higher order transition operator must be introduced
\begin{eqnarray}
{\cal S}_{\alpha\alpha',\beta\beta'}
&=&
\sum_{I,J}{\cal B}_{IJ}^{\rm S}
\left[
\frac{1}{2} 
\frac{\partial (E_{\alpha}+E_{\alpha'})}{\partial S_I} d^{J}_{\alpha\beta}\delta_{\alpha'\beta'}
\right.
\nonumber\\
&+&\frac{1}{2} \frac{\partial (E_{\alpha}+E_{\alpha'})}{\partial S_I} d^{J*}_{\alpha'\beta'}\delta_{\alpha\beta}
\nonumber\\
&-&\frac{1}{2} \Delta E_{\alpha\sigma}d^{I}_{\alpha\sigma} d^{J}_{\sigma\beta}\delta_{\alpha'\beta'}
-\frac{1}{2} \Delta E_{\alpha\beta}d^{I}_{\alpha\beta} d^{J*}_{\alpha'\beta'}
\nonumber\\
&-& \frac{1}{2}\Delta E_{\alpha'\sigma'}d^{I*}_{\alpha'\sigma'}d^{J*}_{\sigma'\beta'}\delta_{\alpha\beta}
\nonumber\\
&-&\left.\frac{1}{2}\Delta E_{\alpha'\beta'}d^{I*}_{\alpha'\beta'}d^{J}_{\alpha\beta}
\right]
\;.
\label{eq:Sspin}
\end{eqnarray}
The operator in Eq.~(\ref{eq:Sspin}) is identically zero for canonical conjugate variables.

The general equation of motion~(\ref{eq:qcsd-dyna}) is difficult to integrate.
However, one can consider its weak-coupling limit upon taking the adiabatic limit
of the operators in Eqs.~(\ref{eq:Jspin}) and~(\ref{eq:Sspin}).
This is performed considering the off-diagonal elements of $d_{\alpha\alpha'}$, which
couples different adiabatic energy surfaces, to be negligible.
In such a limit, one obtains
\begin{eqnarray}
J_{\alpha\alpha',\beta\beta'}^{\rm ad}
&=&
- \sum_{I,J}{\cal B}_{IJ}^{\rm S}\frac{\partial H_{\rm SB}}{\partial S_J} 
\left( d^{I}_{\alpha\alpha} + d^{I*}_{\alpha'\alpha'}
\right)\delta_{\alpha\beta}\delta_{\beta'\alpha'}
\nonumber\\
&=&
- i\sum_{I,J}{\cal B}_{IJ}^{\rm S}\frac{\partial H_{\rm SB}}{\partial S_J} 
\left( \phi^{I}_{\alpha\alpha} - \phi^{I}_{\alpha'\alpha'}
\right)\delta_{\alpha\beta}\delta_{\beta'\alpha'}
\nonumber\\
\label{eq:Jspin-ad}
\end{eqnarray}
where using the fact that $d_{\alpha\alpha}^I$ is purely imaginary
a phase
\begin{equation}
\phi^I_{\alpha\alpha}=-id_{\alpha\alpha}^I
\label{eq:geo-phase}
\end{equation}
can be introduced.
In a similar way one can take the adiabatic limit
of the ${\cal S}_{\alpha\alpha',\beta\beta'}$
in Eq.~(\ref{eq:Sspin})
\begin{eqnarray}
{\cal S}_{\alpha\alpha',\beta\beta'}^{\rm ad}
&=&
-\frac{i}{2} \sum_{I,J}{\cal B}_{IJ}^{\rm S}
\frac{\partial (E_{\alpha}+E_{\alpha'})}{\partial S_J} 
\left(\phi^{I}_{\alpha\alpha} - \phi^{I}_{\alpha'\alpha'}\right)
\delta_{\alpha\alpha}\delta_{\alpha'\alpha'}
\nonumber\\
\label{eq:Sspin-ad}
\end{eqnarray}
Hence, the weak-coupling (adiabatic) equation of motion reads
\begin{eqnarray}
\frac{\partial}{\partial t}\rho_{\alpha\alpha'}
&=&\left[-i\omega_{\alpha\alpha'}
-i \sum_{I,J}B_{IJ} \frac{\partial H^{\rm S}_{\alpha\alpha'}}{\partial S_J} 
\left(\phi^{I}_{\alpha\alpha} - \phi^{I}_{\alpha'\alpha'}\right)
\right.\nonumber\\
&-&\left.
\sum_{K,L}B_{KL}\frac{\partial H_{\alpha\alpha'}^{\rm S}}{\partial S_L} \frac{\partial}{\partial S_K}
\right]
\rho_{\alpha\alpha'}
\label{eq:qcsd-dyna-ad}
\end{eqnarray}
In Eq.~(\ref{eq:qcsd-dyna-ad}) one can note the presence of two
phase terms.
While the phase $\omega_{\alpha\alpha'}$ is dynamical,
the phase $\phi_{\alpha\alpha}$ is of a geometric origin
and it is analogous to the famous Berry phase~\cite{berry,qphases,mead}.
It is predicted by Eq.~(\ref{eq:pWS}) to be present also for open paths~\cite{pati}
of the classical environment and it is 
off-diagonal in nature~\cite{filipp,englman,manini}
since it can only be different
from zero for the off-diagonal elements of $\rho_{\alpha\alpha'}$.
It easy to see that the geometric phase $\phi_{\alpha\alpha}$
can be obtained without invoking the adiabatic limit
upon selecting the diagonal elements $d_{\alpha\alpha'}^I$
in Eq.~(\ref{eq:qcsd-dyna}).
Hence, it is remarkable that the geometric phase $\phi_{\alpha\alpha}$
is predicted also for non-adiabatic dynamics.
The prediction of the geometric phase in Eq.~(\ref{eq:geo-phase})
is the second important result of this work.

In the absence of explicit time-dependences, Eq.~(\ref{eq:qcsd-dyna-ad})
can be rewritten as
\begin{eqnarray}
\partial_t\rho_{\alpha\alpha'}
&=&\left[-i\omega_{\alpha\alpha'}
-\left( \langle\alpha,S|\frac{d}{dt}|\alpha,S\rangle -\langle\alpha',S|\frac{d}{dt}|\alpha',S\rangle \right)
\right.\nonumber\\
&-&\left.
\sum_{I,J}{\cal B}_{IJ}^{\rm S}
\frac{\partial H_{\alpha\alpha'}^{\rm S}}{\partial S_J} \frac{\partial}{\partial S_I}
\right]
\rho_{\alpha\alpha'}
\label{eq:qcsd-dyna-ad2}
\end{eqnarray}
Using Dyson identity, this can be written in propagator form as
\begin{eqnarray}
\rho_{\alpha\alpha'}(t)
&=&\exp\left[-i\int_{t_0}^t dt'\omega_{\alpha\alpha'}(t')\right]
\nonumber\\
&\times&\exp\left[
-\int_{t_0}^t dt'\left( \langle\alpha,S|\frac{d}{dt'}|\alpha,S\rangle
 -\langle\alpha',S|\frac{d}{dt'}|\alpha',S\rangle \right)\right]
\nonumber\\
&\times&
\exp\left[ -(t-t_0)\sum_{I,J}{\cal B}_{IJ}^{\rm S}
\frac{\partial H_{\alpha\alpha'}^{\rm S}}{\partial S_J} 
\frac{\partial}{\partial S_I}
\right]
\rho_{\alpha\alpha'}(t_0)
\label{eq:qcsd-dyna-ad-prop}
\end{eqnarray}
Equation~(\ref{eq:qcsd-dyna-ad-prop}) is the
adiabatic approximation of the quantum-classical Liouville
equation for spin baths, whose general form is given in Eq.~(\ref{eq:qcsd-dyna}).
The propagator form of Eq.~(\ref{eq:qcsd-dyna-ad-prop}) is suitable
to the development of an effective numerical scheme:
as in the case of baths of canonically conjugate
phase space coordinates, the evolution of the quantum system
can be represented in term of the propagation of classical spin
trajectories over coupled adiabatic energy surfaces.
Such a scheme is complementary to those
based on the solutions of master equations~\cite{petruccione,ottinger-qm,ottinger-qm2,ottinger-qm3,ottinger-qm4,ottinger-qm5}.
However, it does not require to approximate
in any form the memory effects of the bath since its degrees
of freedom are described explicitly, in the spirit of molecular
dynamics simulations~\cite{allentildesley,frenkelsmit}. 
From this point of view, the approach
presented here provides a non-Markovian route to
the simulation of quantum effects in classical spin baths.
It is expected that the theory and numerical schemes presented here
have a wide applicability~\cite{prokofev,wang-shao,schnack,schroeder,schroeder2}.

This work is based upon research supported by
the National Research Foundation of South Africa.


\end{document}